# Two Memorable Dates in Seismology

Anatol Guglielmi[1], Boris Klain[2], Alexey Zavyalov[1], and Oleg Zotov[1,2]

[1] Schmidt Institute of Physics of the Earth, Russian Academy of Sciences; Bol'shaya Gruzinskaya str., 10, bld. 1, Moscow, 123242 Russia; guglielmi@mail.ru (A.G.), zavyalov@ifz.ru (A.Z.); ozotov@inbox.ru (O.Z.)

[2] Borok Geophysical Observatory of Schmidt Institute of Physics of the Earth, Russian Academy of Sciences; klb314@mail.ru (B.K.)

**Abstract:** 130 years ago, Omori formulated the first law of earthquake physics. The essence of the law is that the frequency of aftershocks decreases hyperbolically over time. 100 years ago, Hirano doubted the universality of Omori's law and proposed a power law for the evolution of aftershocks. Our paper is devoted to these two outstanding events, which played a significant role in the development of seismology. The paper also contains a brief summary of the modern approach to the construction of a phenomenological theory of aftershocks. Within the framework of the phenomenological approach, the epoch of harmonic evolution of the source, "cooling down" after the main shock, the bifurcation of the source, the cumulative effect of a round-the-world echo, the modulation of global seismicity by spheroidal oscillations of the Earth, mirror triads, migration of aftershocks, and a number of other previously unknown geophysical phenomena were discovered. It is emphasized that the phenomenological theory has been formed in recent years on the basis of a thorough analysis of the fundamental research of the pioneers who laid the foundations of modern seismology.

## Introduction

100 years ago, Hirano proposed a power law for the evolution of aftershocks from a strong earthquake [1], having doubted the universality of the hyperbolic law proclaimed by Omori 130 years ago [2]. Before describing these two outstanding events that played a significant role in the development of seismology, we will briefly outline the modern approach to constructing a phenomenological theory of aftershocks [3].

Let us define the earthquake source as a rock mass inside the convex shell of the aftershock envelope. Let us introduce the phenomenological parameter $\sigma(t)$, which generally characterizes the state of the source at time $t$. We will call the quantity $\sigma$ the the *source deactivation coefficient*. We will assume that the $\sigma$ value indicates the rate at which the source loses its ability to excite aftershocks. Let us replace $t$ with the so-called proper time as follows:

$$\tau = \int_0^t \sigma(t')dt'. \tag{1}$$

Let's try to find the equation for the evolution of the source in the form of a differential equation $dn/d\tau + f[n(\tau)] = 0$, which describes the change in the frequency of aftershocks $n$ over time. For obvious reasons, we will not be able to derive the equation of evolution from first principles. However, the equation can be guessed, and then the guess can be tested in an experiment.

It is known that the frequency of aftershocks on average decreases over time more slowly than exponentially [4]. Therefore, the choice of a linear equation does not suit us. Let's try to use the simplest nonlinear equation

$$\frac{dn}{d\tau} + n^2 = 0. \tag{2}$$

The choice [5] of the equation with quadratic nonlinearity turned out to be extremely successful. First, the evolution equation (2) allows for interesting generalizations [3, 6]. Secondly, and this is the main thing, it allows a new approach to the processing and analysis of aftershocks.

Let us rewrite (2) in the form

$$\frac{dn}{dt} + \sigma n^2 = 0, \tag{3}$$

and pose the inverse source problem for equation (3). The essence of the inverse problem is to calculate the deactivation coefficient $\sigma(t)$ from the aftershock frequency $n(t)$ known from observations [7].

The procedure for calculating $\sigma(t)$ is simple. We introduce an auxiliary function $g(t) = n^{-1}(t) - n_0^{-1}$, where $n_0 = n(0)$ We rewrite (3) in the form of an integral relation

$$\int_0^t \sigma(t')dt' = g(t). \tag{4}$$

The formal solution to the task is obvious: $\sigma = dg/dt$. However, this solution is unstable, since the inverse problem is formulated incorrectly, as is usually the case when setting inverse problems in geophysics. The incorrectness is associated with rapid fluctuations of the function $n(t)$. Regularization in this case comes down to smoothing the auxiliary function. As a result we have

$$\sigma = \frac{d}{dt}\langle g \rangle, \tag{5}$$

where the angle brackets indicate the smoothing procedure.

Within the framework of the phenomenological approach, a number of interesting geophysical phenomena have been discovered: harmonic evolution of the source [8], source bifurcation [9], cumulative effect of round-the-world echo [10, 11], modulation of global seismicity by free vibrations of the Earth [12], mirror triads and solitary strong earthquakes [13], migration of aftershocks [14].

We emphasize that, following Hugo, we recognize that the new is based on the previous. And we fully understand that the phenomenological theory of aftershocks [3] was formed under the strong influence of the works of the founders of seismology [1, 2].

## Omori (1894)

It is best to start describing the outstanding events of a century ago from afar. In 1850, John Milne was born in Liverpool, who in the future was to become a famous geophysicist, one of the founders of modern seismology [15]. He was educated in London, worked as a mining engineer in Newfoundland, as a geologist in the Sinai Peninsula, and from 1875 to 1895 worked in Tokyo at the invitation of the government of the Japanese Empire. It is curious that in search of adventure, Milne traveled to Tokyo for three months, mostly by land (via Siberia). In 1880, he created a horizontal pendulum seismograph - the first easy-to-use and sufficiently sensitive instrument for recording earthquakes.

In 1887 Milne was elected a member of the Royal Society. He manages to convince the Society to allocate funds for the creation of a world network of seismic stations equipped with his instruments. Japan highly appreciated Milne's services to the country and the world. Emperor Meiji awarded him the Order of the Rising Sun and awarded him a lifetime pension of 1,000 yen. The University of Tokyo elected him honorary professor.

Fusakichi Omori was a student of John Milne. He enjoyed the encouraging support of his teacher, as did all young Japanese seismologists of that time. On October 28, 1891, the earthquake with a magnitude of $M = 8$ occurred. Milne's seismographs recorded numerous aftershocks. Analysis of these aftershocks allowed Omori to formulate the law that bears his name [2]. It is worth mentioning that he was 26 years old at the time.

Omori's Law states that after a strong earthquake, the frequency of aftershocks, i.e. tremors following the main shock, on average, decreases hyperbolically over time:

$$n(t) = \frac{k}{c+t}. \qquad (6)$$

Here $k > 0$, $c > 0$, $t \geq 0$. This was the first law of earthquake physics, if we keep in mind the chronological sequence of outstanding discoveries in seismology.

Intuition tells us that the Omori formula (6) has the characteristics of a fundamental law of nature. Let's explain what was said. Firstly, (6) can be rewritten in the equivalent form $n = k/t$, where $k > 0$, $t > 0$. In fact, contrary to popular belief, the parameter in (6) is free due to the homogeneity of the flow of time $t$. In other words, Omori's law is one-parameter: the value of $k$ generally characterizes the source of an earthquake over the entire history of the evolution of aftershocks.

Secondly, according to Omori $dn/dt \propto -1/t^2$, which immediately causes us to associate with the nonlinear equation for the evolution of aftershocks (2). It becomes clear that Omori's law, if it is true, is a very specific solution of the more general law (2). Let's find the state of the source in which Omori's law (6) directly follows from the fundamental phenomenological equation (2).

The general solution of equation (2) is

$$n(\tau) = \frac{n_0}{1 + n_0 \tau}. \qquad (7)$$

Taking (1) into account, it is easy to see that Omori's law (6) follows from (7) under the condition $\sigma = const$. In this case $k = 1/\sigma$, $c = 1/n_0\sigma$. In other words, Omori's law is satisfied provided that the deactivation coefficient remains unchanged throughout the entire history of aftershock excitation.

The found condition for the applicability of the Omori law is tough. Is it carried out experimentally? To answer the question, we compiled the Aftershock Atlas [8], which presents solutions to the inverse source problem. The characteristic size of the source was determined according to the method [16].

Analysis of solutions for several dozen events indicates that in all cases, at the first stage of the evolution of aftershocks, immediately after the main shock, the source deactivation coefficient remains unchanged within the limits of measurement accuracy. We call the time interval during which $\sigma = const$ the Epoch of Hyperbolic Evolution (EHE), or Omori epoch. The duration of EHE varies from case to case from a few days to several months. Figure 1 shows a case of a fairly long EHE, almost equal to 100 days. There is a tendency for $\sigma$ to decrease with increasing magnitude of the main shock [17].

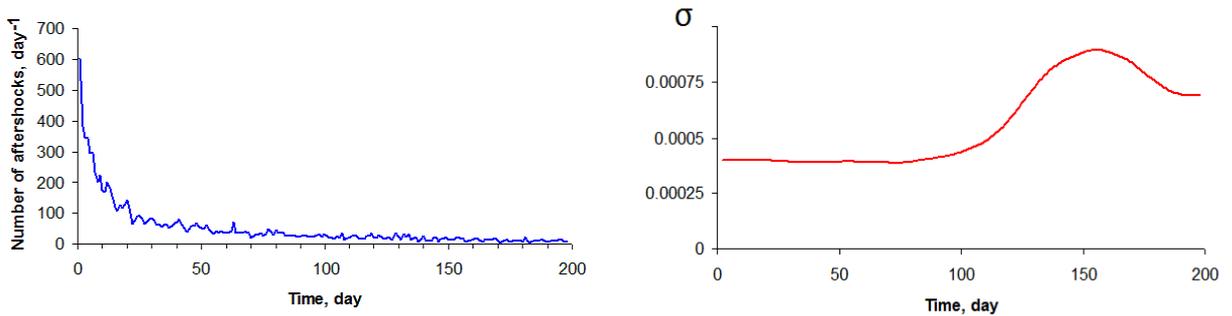

**Fig. 1**. Frequency of aftershocks (left) and source deactivation factor (right) after the earthquake with magnitude M = 6.7 at a depth of H=18.2 km in Southern California on 1994.01.17 (Northridge earthquake).

The discovery of the Omori epoch is extremely significant. At the end of the epoch, the state of the source changes. Let's introduce parameter $\theta = d\sigma/dt$. In the Omori epoch, $\theta = 0$. The transition to a new state is indicated by a sharp jump in parameter $\theta$ (Figure 2). This suggests that the end of the Omori epoch is accompanied by a bifurcation of the earthquake source.

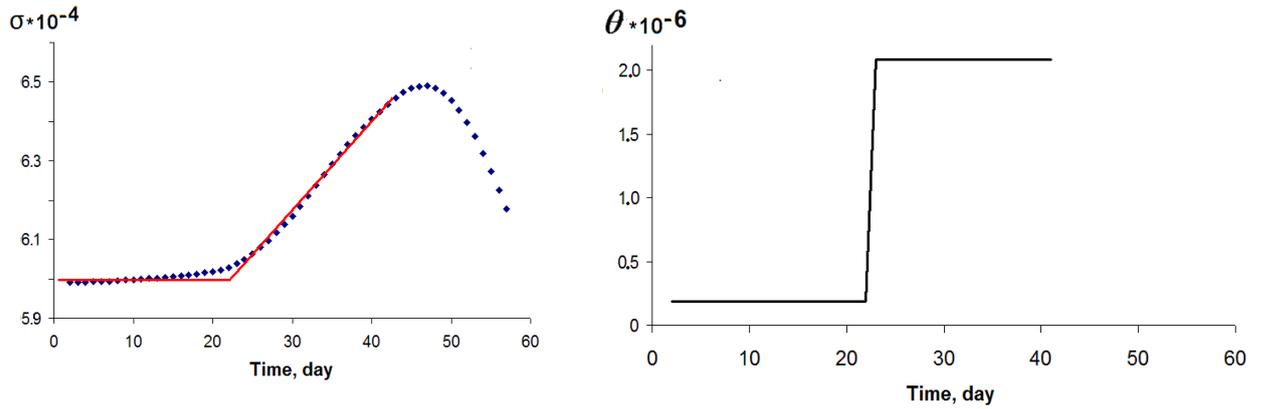

**Fig. 2.** Piecewise linear approximation of the deactivation coefficient (left) and its time derivative (right). The event occurred on November 23, 1984 in Northern California. The magnitude of the main impact is $M = 6$, the depth of the hypocenter is 10 km.

A convincing illustration of this is Figure 3. It shows the result of processing and analyzing eight events. The transition from one mode of source deactivation to a qualitatively different mode is abrupt in the sense that its duration is much less than the duration of the Omori epoch.

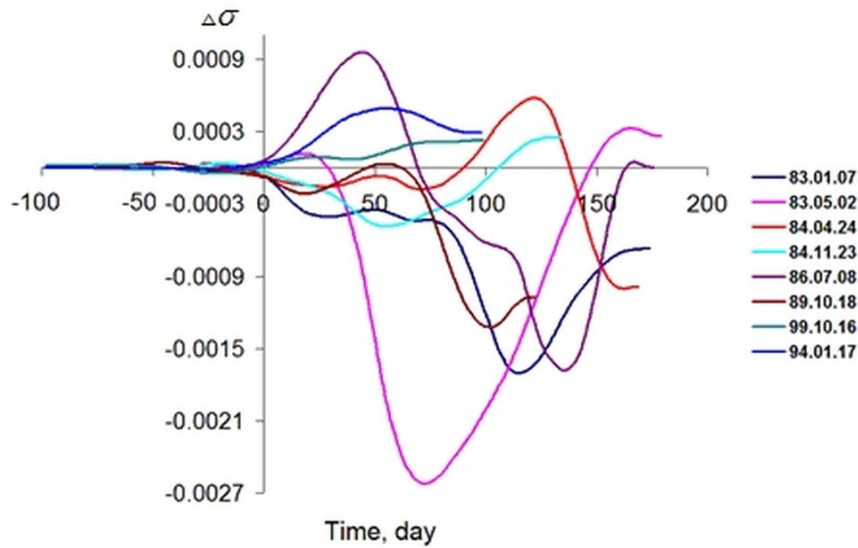

**Fig. 3**. Summary bifurcation diagram indicating a change in the deactivation mode of the earthquake source after the end of the Omori epoch (see text). The vertical axis shows the deactivation factor minus the value of this factor at the beginning of the Omori epoch. The critical point is aligned with zero point in time.

Let us summarize the preliminary results of studying the evolution of aftershocks by solving the inverse problem:
1. The Omori formula (6) has the status of a law that is fulfilled with good accuracy in a real geophysical process.

2. The applicability of Omori's law is strictly limited to the epoch of harmonic evolution of the earthquake source.

3. At the end of epoch of harmonic evolution, the source experiences a bifurcation.

To conclude the section devoted to Fusakichi Omori, let us return to formulas (6) and (7). The law of aftershocks (7) formally differs from law (6) only in that in it time $t$ is replaced by proper time $\tau$. The concept of proper time and the corresponding term were introduced into physics only at the beginning of the 20th century. It would seem that Omori simply did not have the opportunity to use the concept of proper time in order to give formula (6) the flexibility and physical meaning that our formula (7) has. In this regard, we want to clarify that the source proper time (1) is not directly related to the relativistic proper time. Formulas of the form (1) express the uneven functioning of any dynamic system immersed in a nonstationary environment or located in nonstationary external force fields. In our case, formula (1) takes into account the non-stationary state of the source that experienced the main shock of the earthquake.

**Hirano (1924)**

100 years ago, an event occurred that determined for a long time the style of studying the evolution of aftershocks: Hirano published an article [1], in which he proposed replacing Omori's law (6) with a power law of the form

$$n(t) = \frac{k}{(c+t)^p}. \qquad (8)$$

The years went by. In 1938, the famous mathematician and geophysicist Jeffreys drew attention to Hirano's formula [19]. Another couple of decades passed. And so, thanks to the efforts of Utsu [20–22], formula (8) finally attracted the close attention of seismologists. Since then, formula (8) has become firmly established in the practice of aftershock research. It was found that the exponent is on average $p = 1.1.$, but varies from case to case within wide limits (approximately from 0.7 to 1.5).

Formula (8), like Omori formula (6), is one-parameter. In fact, we have already spoken about the parameter $c$ in connection with formula (6). As for the parameter $k$, it does not have any physical meaning for Hirano, since it does not have a fixed dimension. The dimension $k$ depends on the value $p$, and this is not accepted in physics. Thus, the Hirano-Utsu formula (8) does not have the status of a geophysical law, but nothing prevents us from considering it as a completely acceptable fitting formula for approximating observational data.

We would like, with the greatest respect to the discoverer, to express directly our opinion that Hirano's search for a correct aftershock formula, different from the Omori formula, was justified, since the Omori formula (6) is not universal. Another thing is that the choice of Hiroano's formula (8) turned out to be unsuccessful. However, it would be a mistake to think that we do not appreciate Hirano's creative efforts. After all, we always take risks, since our main method of searching for scientific truth is trial and error. One way or another, the long search in this direction, begun by Milne and Omori in the century before last, continued by Hirano and Utsu in the last century, has finally been successfully completed. The universal one-parameter formula for the evolution of aftershocks (7) has been found. It differs from the Omori formula (6) in that it takes into account the difference between the proper time $\tau$ and the world time $t$.

And finally, one more historical observation. It is quite clear that formulas (6)–(8) are not only pictograms that schematically reflect one or another seismologist's idea of the direction of the evolution of aftershocks. These formulas are primarily and mainly tools for processing and analyzing observations in order to calculate the phenomenological parameters of the earthquake source. Omori formula (6) allows you to calculate the parameter $k$. The Hirano-Utsu formula (8) allows you to calculate parameter $p$. Formula (7), by which we calculate the source deactivation factor $\sigma$, turned out to be logically more justified and much more effective.

## Discussion

We believe that the arguments in favor of the phenomenological theory of aftershocks are stronger and more convincing than the arguments in favor of the old theory based on the Hirano-Utsu formula. One of the arguments in favor of a differential evolution equation containing quadratic nonlinearity is that the form of the equation suggests natural and very effective generalizations.

An obvious generalization of (2) is the inhomogeneous differential equation

$$\frac{dn}{d\tau} + n^2 = f(t). \qquad (9)$$

The free term $f(t)$ can be used to formally describe the triggers that induce aftershocks. Here we remember two endogenous triggers, one of which is pulsed ($f(t) \propto \delta(t)$), and the second sinusoidal ($f(t) \propto \sin(\omega t)$). The idea is that the main shock of an earthquake excites two non-trivial triggers, namely, a round-the-world seismic echo and free vibrations of the Earth, which can affect the dynamics of the "cooling" earthquake source.

With some stretch, the theory of the two triggers we have indicated could be called semi-phenomenological, since at least the echo delay time and the period of free oscillations of the Earth are calculated within the framework of the fundamental theory of elastic oscillations of the Earth. In the future, we expect that the phenomenological parameters and, above all, the source deactivation coefficient will be calculated on the basis of first principles, just as the gas temperature is calculated within the framework of kinetic theory. But the theory of aftershocks is still very, very far from such an ideal.

Generally speaking, generalization (9) is quite trivial. And it only designates exogenous and endogenous triggers, helps to classify them and the like. A nontrivial generalization of equation (2) is the logistic equation for aftershocks [23]

$$\frac{dn}{dt} = n(\gamma - \sigma n) \qquad (10)$$

This equation is well known in biology, chemistry, sociology and other sciences. Here $\gamma$ is the second phenomenological parameter of our theory. Its approximate estimate can be made using the formula $\gamma = \sigma n_\infty$, where $n_\infty$ is the frequency of tremors in the background seismicity regime, $n_\infty \ll n_0$.

It should be noted that we were led to the logistic equation (9) by the research of Faraoni [24], who saw a remarkable analogy between the aftershocks equation (2) and one of Friedman's equations, which describes the expanding Universe. Equation (9) describes the evolution of aftershocks under conditions of limited influx of free energy and negative entropy into the source. An interesting perspective opens up when considering the stochastic analogue of the logistic equation for aftershocks.

We started with the simplest nonlinear evolution equation (2), carefully substantiated its applicability for describing aftershocks, and with great care made minimal generalizations reflecting certain features of the source dynamics, while trying not to destroy the basis of our phenomenological theory. Now we want, with all the necessary reservations, to make a radical generalization and consider the source as a spatially extended object [3, 7, 14, 25]. To do this, we replace the frequency of aftershocks $n(t)$ with the distribution density of aftershocks $n(t, \mathbf{x})$, where $\mathbf{x} = (x, y, z)$ are the coordinates of the hypocenter. If the distribution of epicenters is analyzed, then $\mathbf{x} = (x, y)$ should be set and the origin of the coordinate system should be aligned with the epicenter of the main shock. Finally, sometimes the epicenters of aftershocks are located predominantly in the vicinity of a long fault in the earth's crust. In this case, a simplified description of the distribution of aftershocks using the function $n(t,x)$ is acceptable, with the distance $x$ measured along the fault.

To describe space-time dynamics, we need a partial differential equation. Trying to make minimal changes to the logistic equation for the evolution of aftershocks, we add to its right side an additional term of the form $\hat{L}[n(t,x)]$, in which $\hat{L}$ is a linear differential operator. Let's take $\hat{L} = \partial^2/\partial x^2$ and get the equation for the spatiotemporal evolution of aftershocks

$$\frac{\partial n}{\partial t} = n(\gamma - \sigma n) + D\frac{\partial^2 n}{\partial x^2}. \qquad (11)$$

Here $D$ is the third phenomenological parameter of our theory.

The nonlinear diffusion equation (11) is known in mathematics as the Kolmogorov–Petrovsky–Piskunov equation [26]. It has solutions in the form of stationary waves propagating with velocity $V \sim \sqrt{D\gamma}$. Knowing $V$ and $\gamma$, we can estimate the diffusion coefficient $D$.

Along this path, the complex task of experimentally searching for directional migration of aftershocks was formed. The problem was first successfully solved in [14] (see also [3, 25]). The difficulty is this. We want to consider the source as a kind of track detector. We will monitor the movement of aftershocks and try to estimate the velocity of movement. But we talk about the movement of aftershocks only figuratively. Each discontinuity that excites an aftershock occurs locally. In fact, the localzation of the sequence of ruptures moves underground.

Leaving aside subtle conceptual issues, we want to make quite convincing only the very fact of the slow systematic removal of aftershocks from the epicenter of the main shock. To do this, we need a large amount of observational data. We extracted information on 138 events from the USGS/NEIC catalog for the years 1973–2019. Main shocks with magnitude $M > 7.5$ were selected. We processed 34627 aftershocks that were observed within 100 days after the main shocks. The processing consisted, firstly, in synchronizing events according to the time of the main shock. Secondly, a non-standard technique for ordering aftershocks was used, proposed by one of the authors (Oleg Zotov): aftershocks were combined into clusters, each of which was determined by the aftershock serial number. Clustering smooths out sharp differences in the deactivation coefficient, on which the rate of flow of proper time in the source depends. Finally, the average distance $r = \sqrt{x^2 + y^2}$ between the aftershocks and the main shock was calculated for each cluster. The result is shown in Figure 4. Red line

$$r[\text{km}] = 0.08 j + 67.9 \qquad (12)$$

approximates experimental points. Here $j$ is the cluster number. The correlation coefficient is 0.78. Undoubtedly, aftershocks move away from the main shock over time.

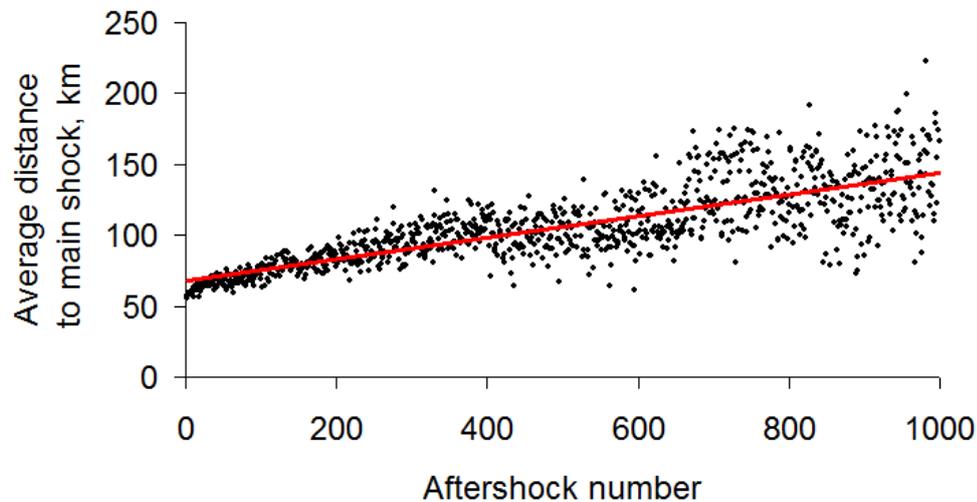

**Fig. 4.** Removal of aftershock epicenters from the main shock epicenter over time during the evolution of the earthquake source.

Recognizing that the flow of aftershocks moves away from the main shock is a big step toward understanding earthquake mechanics. Here the key results have not yet been obtained. We still have to learn how to calculate the flow of Umoff's mechanical energy in the source of earthquake. We need to understand the mechanism of formation of the Umoff flow velocity in a rock mass. Intuition tells us that the divergent flow of aftershock energy may correspond to a converging flow of foreshock energy, and we still have to test this hypothesis experimentally.

So, 130 years have passed since the search for the law of aftershock evolution began. The process turned out to be complex and contradictory. And, of course, one cannot say that the process has already been completed. But we can make a number of statements quite definitely:
- the law of aftershock evolution has the form of a differential equation containing quadratic nonlinearity,
- Omori's law is fulfilled, but has a time-limited range of applicability,

– at the end of the Omori epoch, the state of the source as a dynamic system changes sharply, which probably indicates the phenomenon of bifurcation,
– there is a slow removal of the epicenters of aftershocks from the epicenter of the main shock.

**Conclusion**

So, at the end of the century before last in Japan, the birth of modern seismology occurred due to the fact that at this time and in this place the urgent need of society, state support and human genius miraculously united. We recalled the background and briefly told the story of the discovery of the first law of earthquake physics.

At the beginning of the last century, Hirano drew attention to the fact that the Omori formula does not always satisfactorily describe the flow of aftershocks, and proposed his own version of the law of evolution. Several decades later, Utsu carried out a series of studies using Hirano's formula to process and analyze observations.

The results obtained by Utsu aroused keen interest, were continued, and generated an extensive literature. A thorough study of research carried out using the Utsu method led us to the idea of constructing a phenomenological theory of aftershocks, starting from the simplest differential evolution equation with quadratic nonlinearity.

Within the framework of the phenomenological approach to the physics of aftershocks, it was possible to obtain a number of previously unknown results. In conclusion, we list some of them again:

1. It has been established that Omori's law is true, but only for a limited period of time after the main shock,
2. The phenomenon of bifurcation at the end of the epoch of harmonic evolution of the source was discovered,
3. The Hirano-Utsu formula has been analyzed and shown that it can be used as a fitting formula, but does not have the status of a geophysical law,
4. The cumulative effect of round-the-world seismic echo was predicted and discovered,
5. Modulation of global seismicity by spheroidal oscillations of the Earth was discovered,
6. The phenomenon of aftershock migration was discovered, and an interpretation of migration was proposed within the framework of the Kolmogorov-Petrovsky-Piskunov theory of nonlinear diffusion waves.